\documentclass[twocolumn,showkeys,showpacs,nofootinbib,aps,epsfig]{revtex4}

\usepackage{graphicx}
\def\bea{\begin{eqnarray}}
\def\eea{\end{eqnarray}}

\newcommand{\na}{\nabla}
\def\beq{\begin{equation}}
\def\eeq{\end{equation}}

\def\pa{\partial}

\def\na{\nabla}
\newbox\pippobox

\def\bea{\begin{eqnarray}}
\def\eea{\end{eqnarray}}
\newcommand{\nn}{\nonumber}
\def\beq{\begin{equation}}
\def\eeq{\end{equation}}

\def\pa{\partial}

\def\na{\nabla}

\def\M{\mu}
\def\N{\nu}
\def\vp{\varphi}
\def\beq{\begin{equation}}
\def\eeq{\end{equation}}

\def\pa{\partial}

\newbox\pippobox

\begin{document}

\title{Conformally-coupled dark spinor and FRW universe}
\author{Joohan Lee}
\address{Department of Physics,
University of Seoul, Seoul 130-743 Korea}
\email{joohan@kerr.uos.ac.kr}

\author{Tae Hoon Lee}
\address{Department of Physics and Institute of Natural
Sciences, Soongsil University, Seoul 156-743 Korea}
\email{thlee@ssu.ac.kr}

\author{Phillial Oh}
\address{Department of Physics
and Institute of Basic Science, Sungkyunkwan University, Suwon
440-746 Korea}
 \email{ploh@skku.edu}

\date{\today}

\begin{abstract}
We study conformal coupling of dark spinor fields to gravity and
calculate the energy density and the pressure of the spinor in FRW
spacetime. We consider the renormalizable potential of the spinor
field. In the cases where the field is proportional to some power of
the cosmic scale factor $a(t)$, we determine the Hubble parameter as
a function of the scale factor and find analytic solutions for
$a(t)$ when the spinor field matter dilutes as the universe expands.
We discuss the possibility that both matter- and dark
energy-dominated eras of our universe can be described by the dark
spinor.
\end{abstract}



\maketitle

\section{Intoduction}

Recently various proposals have appeared in the literature to study
the properties of spinor fields in cosmology. Relatively earlier
attempts are the standard spinor fields with a non-linear self
interaction term \cite{Saha}. Although introducing a non-linear
fermionic potential term may cause a problem at the quantum level,
it was treated classically and a number of cosmological issues were
investigated. For instance, the problem of initial singularity, the
issue of isotropization, and their possible role in the late time
acceleration have been studied in this context.

More recently a proposal was put forward to use the so-called ELKO
spinors \cite{ELKO} as a dark matter candidate. One of the most
interesting properties associated with these non-standard spinors is
the fact that their dominant interaction is via the gravitational
field, which is an essential propertiy for any dark matter
candidate. Another interesting feature is that these dark spinors
have a canonical mass dimension one instead of 3/2 for the standard
spinors. This feature makes wider range of perturbatively
renormalizable self interactions possible. Consequently, many
studies on the cosmological aspects of the ELKO spinor model have
followed \cite{his}.

In Ref. \cite{BBMS}, however, it was pointed out that the
construction of ELKO spinors itself implicitly violates Lorentz
invariance. They proposed a non-local but Lorentz invariant version
of the dark spinors. They also pointed out that some crucial errors
were made in the calculation of stress energy tensors in the
previous works on ELKO model. Based upon a careful recalculation
they give a correct expression for the ELKO spinor field and apply
the result to show the existence of de Sitter like solutions.
Dynamical analyses of the ELKO model have followed this work using
the correct form of the stress energy momentum tensor and it was
shown that scaling attractor solutions do not exist in this model
\cite{WS}.

In this paper we study the cosmology with the (ELKO) dark spinor
field treating it as the main part of the matter. For self
interaction we choose the most general renormalizable form of the
potential. Most importantly we include the conformal coupling of the
spinor field with the gravitational field. Although this possibility
was briefly mentioned in Ref. \cite{BBMS}, it was not pursued
further. We will focus on the existence of the cosmological solution
where the spinor field is proportional to the simple power of the
scale factor. Such solutions would be possible only when the
potential terms are chosen appropriately. However even if we
restrict it to the renormalizable type of potential, it turns out
that many different types of solutions are possible.

We fix the notation following Ref. \cite{BBMS} with the metric sign
convention changed and introduce the conformal coupling of dark
spinor field to gravity in Sec. II. Considering the renormalizable
potential of the field, in the cases where the field is proportional
to some power of the cosmic scale factor we study the
Friedmann-Robertson-Walker (FRW) cosmology in Sec. III, and we will
discuss on their physical implications in Sec. IV.

\section{Conformal coupling of dark spinor to gravity}

We study the conformal coupling of dark spinor field to gravity.
Let us consider the action of the form \bea S=&&\int d^{4}x
\sqrt{-g} \big[ ( \frac{1}{2\kappa}-\frac{\beta}{2}\bar{\psi}
\psi) R \label{ein}
\\&& -\frac{1}{2}g^{\M\N}\na_{\M}\bar{\psi}\na_{\N}\psi-V(\bar{\psi}\psi)\big]
+S_{m}, \nn  \eea where $\psi$ is the dark spinor field \cite{ELKO,
BBMS}, $V$ is its potential, and $S_m$ is the action for other
matter. The covariant derivative on a spinor and its dual
$\bar{\psi}$ are given by \beq \na_{\M}\psi \equiv \pa_{\M}\psi
-\Gamma_{\M}{\psi} \eeq and \beq \na_{\M}\bar{\psi}
\equiv\pa_{\M}\bar{\psi}+\bar{\psi}\Gamma_{\M}, \eeq where
$\Gamma_{\M}=\frac{i}{4}\omega_{\M}^{a b} f_{a b}$ with the spin
connection $\omega_{\M}^{a b}=e_{\N}^a(\pa_{\M} e^{\N
b}+\Gamma^{\N}_{\M \rho}e^{\rho b})$, $f_{a b} = -\frac{i}{2}[
\gamma_{a}, \gamma_b]$, and $\gamma_a$-matrices satisfy the Clifford
algebra, $\{\gamma_a, \gamma_b\}=-2\eta_{a b}$, in a locally flat
inertial coordinate with $\eta_{a b}\equiv diag.(-1,1,1,1)$.

Einstein's field equations in the presence of the
conformal coupling become: \beq
(1-\beta\kappa\bar{\psi}\psi)(R_{\M\N}-\frac{1}{2}g_{\M\N}R)
=\kappa(T_{\M\N}^{(e)}+T_{\M\N}^{(m)}+\beta T^{(c)}_{\M\N}),
\label{eins} \eeq where $T_{\M\N}^{(e)}=
\na_{\M}\bar{\psi}\na_{\N}\psi -\frac{1}{2}g_{\M\N}\big(
\na^{\rho}\bar{\psi} \na_{\rho}\psi + 2V \big) + \frac{1}{2}\na_\rho
{J_{\M\N}}^{\rho}$ is the energy-momentum tensor of dark spinor and
this accurate expression containing contributions due to the spin
connection has been obtained in Ref. \cite{BBMS} with $ J{_{\mu
\nu}}^{ \rho} = -\frac{i}{2} [
\nabla_{(\mu}\bar{\psi}{f_{\nu)}}^\rho \psi + \bar{\psi}
{f^{\rho}}_{ ( \mu } \nabla_{ \nu ) }\psi]$. The second term
(proportional to $\beta$) in the left hand side of Eq. (\ref{eins})
and the third term in the right hand side,
$T^{(c)}_{\M\N}=\na_\M\na_\N(\bar{\psi}\psi) - g_{\M\N}
g^{\sigma\rho}\na_\sigma\na_\rho (\bar{\psi}\psi)$, are
contributions from the conformal coupling in the action (\ref{ein}),
$-\frac{\beta}{2}\bar{\psi} \psi R$, and
$\sqrt{-g}~T_{\M\N}^{(m)}=-2{\delta S_m}/{\delta g^{\M\N}}$ with
$\kappa=8\pi G $.

\section{Dark spinor in the FRW cosmolgy}

To study effects of the dark spinor on cosmology in the flat FRW
spacetime with a metric of the form \beq g_{\M\N}=diag.\big(-1,~
a^2(t), a^2(t), a^2(t)\big) \label{rw} \eeq where $a(t)$ is the
scale factor of our universe, we assume \beq \psi(x^\M)=\varphi(t)~
\xi \label{elko} \eeq with a homogeneous real scalar field
$\varphi(t)$ and a constant spinor $\xi$ such that $\bar{\xi} \xi=1$
but $\na_\M \xi \ne 0$ \cite{BBMS}.

We comment that the spinor nature of the dark spinor remains alive in the
real scalar field $\varphi(t) $ defined in Eq. (\ref{elko}) and the ensuing equations
contain the contributions, $3/8 H^2 \varphi^2 + . . .$ in Eqs. (\ref{rhoo}) and (\ref{rho}), from the spin connection which are absent in the case of a genuine scalar field. It was shown that these new contributions of the dark spinor to $\rho_e$ and $p_e$ supply fundamentally different aspects in cosmology \cite{BBMS}.


Applying Einstein's equations (\ref{eins}) in the previous section
to the metric (\ref{rw}), we have two differential equations with
$H=\dot{a}/a$:
\bea &&3(1-\beta\kappa\varphi^2)H^2=\kappa(\rho_e+\rho_{m}), \label{eoms1}\\
&& -(1-\beta\kappa\varphi^2)(2\frac{\ddot{a}}{a}+H^2)
=\kappa(p_e+p_{m}), \label{eoms2} \eea where $\rho_{e}$ is the
energy density of dark spinor, $p_{e}$ is its pressure, and the
energy density and pressure of other matter, $\rho_m$ and $p_m$:
\bea \rho_{e}&&=\frac{\dot{\vp}^2}{2}+V+ \frac{3}{8}H^2\vp^2 +6\beta
H\vp\dot{\vp}, \label{rhoo} \\p_{e}&&=\frac{\dot{\vp}^2}{2}-V-
\frac{3}{8}H^2\vp^2-\frac{1}{4}\dot{H}\vp^2
-\frac{1}{2}{H}\vp\dot{\vp} \nn\\&&
-\beta(4H\vp\dot{\vp}+\pa_t^2\vp^2). \label{rho}\eea
The field equation for dark spinor can be written by means of the
scalar field in Eq. (\ref{elko}): \beq \ddot{\vp}+3H\dot{\vp} +
V,_{\vp}+(\beta{R}-\frac{3}{4}H^2)\vp=0 \label{eoms3}, \eeq where
"$V,_{\vp}$" denotes the derivative of $V$ with respect to $\vp$ and
\beq R=6(\frac{\ddot{a}}{a}+H^2)  \eeq is the scalar curvature.

We take a renormalizable potential of the dark spinor field with
mass dimension one, in the form \beq V=V_0
+\frac{m^2}{2}\vp^2+\frac{\lambda}{4}\vp^4 \label{ren} \eeq with
constants $ V_0$ ($=\Lambda/\kappa$), $m$, and $\lambda$. $\Lambda$
stands for the (positive) cosmological constant \cite{ob}.

\subsection{Ansatz for dark spinor field
and consistency condition for renormalizable potential}

To solve Eqs. (\ref{eoms1}), (\ref{eoms2}), and (\ref{eoms3}), we
assume the following simple form
\footnote{
The merit of the above Ansatz (\ref{ans}) is that its simplicity renders 
the analysis straightforward, and furthermore we speculate that
this hypothesis could be the only one which produces the analytic solutions
of the type, Eqs. (\ref{co}) and (\ref{sinh}), for a realistic cosmological model with the renormalizable potential, Eq. (\ref{ren}).} 
for the scalar field in Eq. (\ref{elko})
 \beq \vp(t)=\vp_0 ~a^n (t) \label{ans} \eeq with constants
$\vp_0$ and $n$.
 When $\rho_m=0$ and $p_m=0$, we can recast the
equations as \beq H^2 (1-c_1 \kappa \vp^2) =\kappa V /3\label{e1}
\eeq with $ c_1= [\frac{1}{8}+\frac{n^2}{6}+\beta(2n+1)]$,
\beq \dot{H }(1-c_2 \kappa \vp^2)=\kappa V /2-\frac{3}{2}H^2 (1-c_3
\kappa  \vp^2) \label{e2} \eeq with $ c_2= [\frac{1}{8} +\beta(n+1)]
$ and $c_3=[1/8+n/6-n^2/6 +\beta/3(3+4n+4n^2)]$, and
\beq V,_{\vp}=\vp[-(n+6\beta)\dot{H}+ c_4 H^2] \label{e3} \eeq with
$c_4=3/4-3n-n^2-12\beta$.

Eqs. (\ref{e1})-(\ref{e3}) give us the consistency condition for the
renormalizable potential $V(\vp)$ in Eq. (\ref{ren}):
 \beq (1-c_1 \kappa \vp^2)[(n+6\beta)\kappa V+2(1-c_2 \kappa
\vp^2)\frac{V,_{\vp}}{\vp}]=\frac{\kappa V}{3}(h_1-h_2 \kappa \vp^2)
\label{con} \eeq with constants $h_1=3/2-3n-2n^2-6\beta $ and
$h_2=(1-2n)(1-4\beta)(3+4n^2+24\beta(1+2n))/16 $. From the above
equation (\ref{con}) we determine the parameters of the potential
(\ref{ren}) as
\beq m^2= \frac{\kappa \Lambda}{12 n} (n+2)(2n+3)(1-2n) =
-2\kappa\Lambda[c_1(n)+c_2(n)] \label{m1} \eeq and \beq \lambda=
4\kappa^2 \Lambda c_1(n) c_2(n) \label{m2}, \eeq
with \bea &&c_1(n)=
\frac{1}{24n}(2n+3)(2n-1)(n+1), \nn\\ &&c_2(n)=
\frac{1}{24n}(2n+3)(2n-1), \nn\\  && c_1(n)+c_2(n)=
\frac{1}{24n}(2n+3)(2n-1)(n+2), \label{m3} \eea where \beq
\beta(n)=\frac{1}{6}-\frac{1}{8n}  \label{beta} \eeq is used.


\subsection{Cosmological solutions}

We restrict our interest to special cases where the cosmic scale
factor $a(t)$ increases but the scalar field $\vp(t) ~( \propto a^n
(t)$ with $n < 0$) decreases with time, and we find useful solutions
to the following equation:

\beq \dot{a}^2 + {\cal V}_{eff} (a)=0 \label{eff} \eeq with \beq
{\cal V}_{eff} (a)=\frac{\Lambda}{3}  [~ c_2 \kappa \vp_0^2 ~
a^{2(n+1)}-a^2], \label{pot} \eeq
which is derived from Eq. (\ref{e1}) and the renormalizable
potential $ V(\vp)=V_0 (1-c_1 \kappa\vp^2)(1-c_2 \kappa\vp^2)$ in
Eq. (\ref{ren}) recast with the help of Eqs. (\ref{m1}) and
(\ref{m2}).

Depending on the sign of $c_2$, two types of solutions are given as
follows.

\subsubsection{$c_2>0$ case}

When $-3/2<n<0$, $c_2>0$ and solving Eq. (\ref{eff}) yields
\beq a(t) \propto {\rm cosh}^{\frac{-1}{n}}
[-n\sqrt{\frac{\Lambda}{3} } (t-t_c)] \label{co} \eeq with a
constant $t_c$.
For $-1<n<0$ the renormalizable potential $V(\vp)$ has stable ground
states, while for $-3/2<n \leq -1$ the potential $V(\vp)$ has no
stable ground state.
In this case, $a \propto  const.  + (t-t_c)^2$ for $t_c \lesssim t$,
and the universe becomes dark energy-dominated for $t_c <<t$
\cite{ob}.

\subsubsection{$c_2<0$ case}

When $ n < -3/2$, $c_2 < 0$ and solving Eq. (\ref{eff}) yields

\beq a(t) \propto {\rm sinh}^{\frac{-1}{n}}
[-n\sqrt{\frac{\Lambda}{3} } t]. \label{sinh} \eeq
For $-2<n<-3/2$ the renormalizable potential $V(\vp)$ has a locally
stable ground state, while for $n \leq -2$ potential $V(\vp)$ has no
stable ground state. Note that $n=-3$ case looks similar to Eq. (33)
of Ref. \cite{Oh}, where stiff fluid- and dark energy-dominated era
of the universe could be described by a special realization of
nonlinear sigma model.

We note that the above equation (\ref{sinh}) leads to
\beq a(t)\sim
t^{-\frac{1}{n}} \label{early}  \eeq for small $t$ and that \beq a
\sim e^{\sqrt{\frac{\Lambda}{3}} t}\eeq for large $t$.
%
%
When we focus on the case where $-2<n<-3/2$ and thus $V(\vp)$ has a
metastable ground state, we might say that
at early times the field $\vp$ acts as dark matter with energy
density and pressure $p_{e}=\omega \rho_{e}$ ($0<\omega<1/3$) and
that at sufficiently late times the field dilutes ($\vp\propto a^n
\rightarrow 0$ with $n<0$) and the cosmological constant
$\Lambda=\kappa V_0$ becomes dominant, leading to an acceleration in
the universe's expansion \cite{ob}, for ${\cal V}_{eff}(a)$ in Eq.
(\ref{pot}) approaches $-\Lambda a^2/3$.

When we calculate the equation of state $\omega=p_{e}/\rho_{e}$ for
the dark spinor field, it is given by
 \beq
\omega =-1
+\frac{n}{6}\frac{(2n-1)(2n+3)\kappa\vp^2}{(6n-\frac{1}{4}(2n-1)(2n+3)\kappa\vp^2)}.
\label{w} \eeq
We notice that $\omega>-1$ for large $t$ in the case $n <-3/2$ and
but $\omega<-1$ in the case $n> -3/2$. ($\omega=-1$ when $n= -3/2$.)
The late-time value of the equation-of-state parameter $\omega$ for
the dark spinor field can reach the parameter value,
$-1.06^{+0.41}_{-0.42}$, reported by using WMAP data \cite{WMAP},
even without any source field carrying negative kinetic energy.

\subsubsection{Massless case ($c_2=0$)}
In the previous two cases {\it 1} and {\it 2}, we exclude one
specific case with $n=-3/2$, where $c_2=0$, $c_1=0$, $V(\vp)=V_0$ as
seen in Eqs. (\ref{m1})-(\ref{m3}), and thus $a \propto
e^{\sqrt{\frac{\Lambda}{3}}t}$.
If the dark spinor had a mass in this case, it could play a role of
dark matter with the scale factor, $a\propto t^{2/3}$ for small $t$,
corresponding to matter-dominated era of our universe.

The other massless case is the $n=-2$ where $H^2\propto const.+
1/a^{4}$, which can describe both radiation- and dark
energy-dominated era of the universe, even though the potential
$V(\vp)$ has no stable ground state.

\section{Summary and Discussions}

Considering dark spinor fields which are non-minimally coupled to
scalar curvature, we have calculated their contribution to the
energy density and pressure in FRW spacetime in Eqs.
(\ref{rhoo})-(\ref{rho}) of Sec. III.
We have used the renormalizable potential with the spinor field
decomposed into a homogeneous real scalar field and a constant
spinor field. In the case where the scalar field is proportional to
some power of the cosmic scale factor, we have found analytic
solutions for the scale factor when the scalar field matter dilutes
as the universe expands. For $-3/2<n<0$ we have got the scale factor
expressed as hyperbolic cosine funtions of the comic time in Eq.
(\ref{co}),
even though only for $-1<n<0$ the potential $V(\vp)$ has stable
ground states.
For $ n < -3/2$ we have obtained the scale factor expressed as
hyperbolic sine funtions of the comic time in Eq. (\ref{sinh}),
even though only for $-2<n<-3/2$ the potential has a locally stable
ground state.
In the latter case, as seen Sec. III B {\it 2}, we have discussed
the possibility that both matter- and dark energy-dominated era of
our universe can be described by the dark spinor.

It is interesting that two specific cases, $n=-3/2$ and $n=-2$, are
not included among the aforementioned cases with $-2<n<-3/2$.
The dark spinor has no self interation in the case of $n=-3/2$, and
the scale factor is determined only by the cosmological constant
$\Lambda$. In the case of $n=-2$ the potential $V(\vp)$ has no
stable ground state. When we see the early time behavior shown in
Eq. (\ref{early}) in the case of $-2<n<-3/2$ where
$11/48<\beta<1/4$, we might think that it corresponds to the
intermediate one between radiation- and matter-dominated era
($0<\omega<1/3$, $\omega = p_{e}/\rho_{e}$). If we add the ordinary
matter contribution in our analysis, then it shall be complicated
but the qualitative features of its results will not be changed.

Even in the case $-1<n<0$ of Sec. III B {\it 1}, we have a non-zero
conformal coupling to gravity, $7/24<\beta$, and we thus notice the
importance of the conformal coupling \cite{BBMS} for the late-time
universe to be described well through the dark spinor.
The late-time value of the parameter $\omega$ in Eq. (\ref{w}) for
the dark spinor converges to a parameter value within the bound
produced by using WMAP data \cite{WMAP}, which has been become
possible in this article without the need for any phantom-like
matter \cite{phantom}, as discussed in Ref \cite{less}.


Our solutions and derived results crucially depend on the
assumptions in Eqs. (\ref{ren}) and (\ref{ans}) of Sec. III.
It would be interesting to check whether relaxing these conditions,
on the renormalizability of $V(\vp)$ or on the configuration of the
scalar field $\vp(t)$, could lead to some more general solutions.

\section*{Acknowledgments}

THL was supported by the Basic Science Research Program through the
National Research Foundation of Korea(NRF) funded by Ministry of
Education, Science and Technology (2012-0003008).\\
PO was supported by the Basic Science Research Program through the
National Research Foundation of Korea(NRF) funded by Ministry of
Education, Science and Technology (2011-0026655).

\end{document}